\documentclass[aps,prl,amsmath,amssymb,floatfix,twocolumn,showpacs]{revtex4-1}
\usepackage{amsmath, dcolumn, bm, amsthm, amsfonts, amssymb, tabularx, slashbox, subfigure}
\usepackage[colorlinks=true,urlcolor=blue,citecolor=blue,linkcolor=blue]{hyperref}
\usepackage{times}
\usepackage{graphicx,graphics,color}
\usepackage{hyperref}
\usepackage{mathdots}
\usepackage{mathrsfs}
\usepackage{changes}

\begin{document}

\newcommand{\vahid}[1]{\textcolor{red}{#1}}
\newcommand{\masoud}[1]{\textcolor{blue}{#1}}

\title{One-dimensional topological metal}

\author{Masoud Bahari and Mir Vahid Hosseini}
\email[Corresponding author: ]{mv.hosseini@znu.ac.ir}
\affiliation{Department of Physics, Faculty of Science, University of Zanjan, Zanjan 45371-38791, Iran}

\date{\today}

\begin{abstract}
We propose a new type of topological states of matter exhibiting topologically nontrivial edge states (ESs) within gapless bulk states (GBSs) protected by both spin rotational and reflection symmetries. A model presenting such states is simply comprised of a one-dimensional reflection symmetric superlattice in the presence of spin-orbit (SO) coupling containing odd number of sublattices per unit cell. We show that the system has a rich phase diagram including a topological metal (TM) phase where nontrivial ESs coexist with nontrivial GBSs at Fermi level. Topologically distinct phases can be reached through subband gap closing-reopening transition depending on the relative strength of inter and intra unit cell SO couplings. Moreover, topological class of the system is AI with an integer topological invariant called $\mathbb{Z}$ index. The stability of TM states is also analyzed against Zeeman magnetic fields and on-site potentials resulting in that the spin rotational symmetry around the lattice direction is a key requirement for the appearance of such states. Also, possible experimental realizations are discussed.
\end{abstract}

\maketitle

{\it Introduction.}---The search for exotic quantum states of matter has attracted a great deal of attention since discovery of topological insulators (TIs) \cite{TI} and topological superconductors (TSs) \cite{Ts} in condensed matter physics. Further investigations have also revealed a novel nontrivial topological states in the so-called Weyl semimetals possessing gapless bulk and Fermi arc surface states \cite{WSMTheo}. In contrast, TIs and TSs have symmetry protected edge states (ESs) inside gapped bulk states. Apart from condensed matter systems, some schemes have been proposed to realize topological phases using cold atoms in optical lattices \cite{TopoCold} and employing light in photonic crystals \cite{TopoPhoto}. Also, the exploration of topological states has been extended even to classical systems \cite{TopoClassic}.

In most of the TIs, the symmetry protected ESs, making relevant requirement for topological quantum computations \cite{TopoCompu}, play dominant role only in a limited certain range of energies. Moreover, due to smallness of energy gap, ESs may also be faded by excitations of bulk states at finite temperature. On the other hand, the coexistence of ESs and bulk states occurs in a narrow energy window in three-dimensional TI candidate materials such as Bi$_2$Se$_3$ and Bi$_2$Te$_3$ \cite{TM}. However, it is intriguing to have a situation in which dominant symmetry protected ESs exist not only in the bandgap but also within the gapless bulk states (GBSs). Therefore, such systems would be in turn served as {\it topological metals} (TMs) \cite{TMCoexsi} even by shifting Fermi level toward conduction or valence bands.
\begin{figure}[t]
  \centering
  \includegraphics[width=8.8cm]{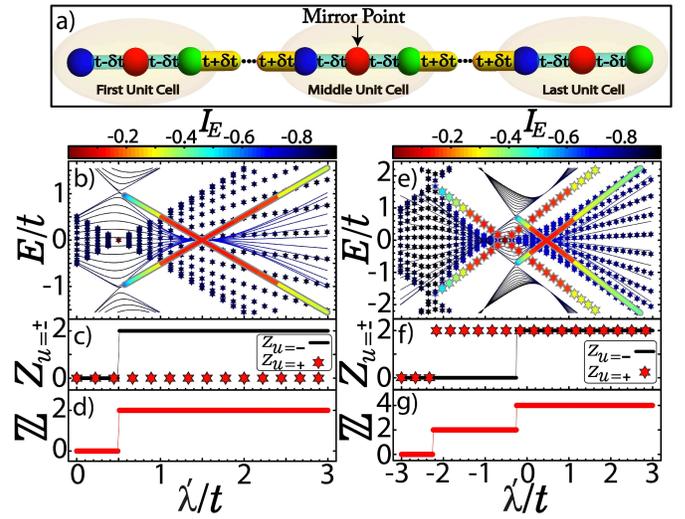}
    \caption{(Color online) (a) Schematic illustration of reflection symmetric superlattice containing three sublattices per unit cell (blue, red and green balls). The mirror point is located at the second sublattice of middle unit cell. The energy spectrum dependence of $\hat{\mathfrak{H}}$ with its IPR on (b) [(e)] $\lambda^\prime$ with $\delta t=-(+)t/2$ and $\lambda_3=-\lambda^\prime$ ($\lambda_3=\lambda^\prime+2.5t$) for 60 unit cells. Solid lines and hexagrams represent the spectra of $\hat{\mathfrak{h}}_{\mathscr{U}=-}$ and $\hat{\mathfrak{h}}_{\mathscr{U}=+}$, respectively. Panels (d) and (g) [(c) and (f)] are the corresponding invariant $\mathbb{Z}$ ($Z_{\mathcal{U}=\pm}$).}
   \label{Fig1}
\end{figure}

\begin{figure}[t]
  \centering
  \includegraphics[width=8.5cm]{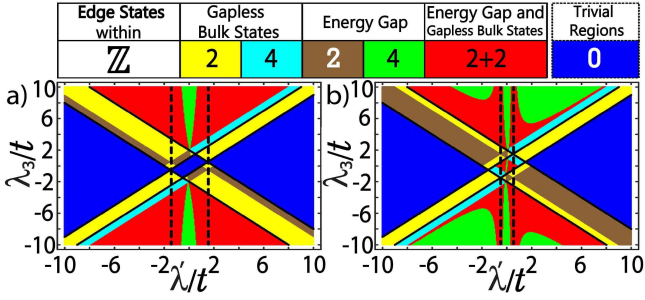}
    \caption{(Color online) Topological phase diagram in the plane $(\lambda^\prime,\lambda_3)$ with $T=3$ and 160 unit cells for (a) $\delta t = -t/2$, (b) $\delta t = +t/2$. The system is a TM at $E=0$ indicated by dashed lines.
    }
   \label{Fig2}
\end{figure}

Several one-dimensional (1D) models have been studied to realize new classes of topological phases \cite{fracCharFer1} concerning both TIs \cite{1DTI} and TSs \cite{1DTS}. These studies stimulate to look for new possibilities for nontrivial topological states mimicking neither TIs nor TSs. So far, however, metallic phase being quasi-degenerate with topologically protected ESs has not been reported to be nontrivial in topology. Hence, it is interesting to develop a minimal and feasible model by which TMs can be emerged easily. In the present Letter, we consider a 1D spin-orbit (SO)-coupled superlattice with odd number of sublattices per unit cell, as shown in Fig. \ref{Fig1}(a), featuring various nontrivial phases [Figs. \ref{Fig2}(a) and \ref{Fig2}(b)]. Surprisingly, we find that topological ESs while keeping their non-triviality can extend into GBSs with increasing SO interaction as shown in Figs. \ref{Fig1}(b) and \ref{Fig1}(e). This results in nontrivial TM phase due to settling SO coupling on odd number of sublattices provided that spin rotational and reflection symmetries are not broken. It should be noted that our findings are valid in general physical grounds, independent of our model parameters, and therefore may be realized in a variety of platforms such as condensed matter systems and quantum Fermi gases.

{\it Model.}---We consider a 1D multipartite SO-coupled superlattice along x-axis with period $T\geq3$ described by the total tight-binding Hamiltonian as
\begin{eqnarray}
\hat{H}=\hat{H}_{t}+\hat{H}_{so},
\label{hamiltonian-real space-total}
\end{eqnarray}
where kinetic ($\hat{H}_{t}$) and SO ($\hat{H}_{so}$) terms are given by $\hat{H}_{t}=\sum_{n,\sigma}\sum_{\alpha}^{T}[t_{\alpha}\hat{c}^\dagger_{\alpha,n,\sigma}\hat{c}_{\alpha+1,n,\sigma}+h.c.]$ and $\hat{H}_{so}=\sum_{n,\sigma}\sum_{\alpha}^{T}[\lambda_{\alpha}\hat{c}^\dagger_{\alpha,n,\sigma}\hat{c}_{\alpha+1,n,-\sigma}+h.c.]$, respectively. The operator $\hat{c}^\dagger_{\alpha,n,\sigma} (\hat{c}_{\alpha,n,\sigma})$  is fermion creation (annihilation) operator of electrons with spin $\sigma = (\uparrow,\downarrow)$ on $\alpha$ sublattice of $n$th unit cell. $t_{\alpha}$ ($\lambda_\alpha$) denotes the hopping (SO coupling) amplitude. Considering periodic boundary conditions and performing fourier transformation, the Hamiltonian (\ref{hamiltonian-real space-total}) can be written in the basis of $\hat{\psi}=(\hat{c}_{1,k,\sigma}, \hat{c}_{2,k,\sigma}, ...,\hat{c}_{T,k,\sigma})^\textit{\textbf{\!T}}$ with $\hat{c}_{\alpha,k,\sigma}=(\hat{c}_{\alpha,k,\uparrow},\hat{c}_{\alpha,k,\downarrow})$ yielding a compact form $\hat{H}=\sum_{k}\hat{\psi}^{\dagger}\hat{\mathcal{H}}(\emph k)\hat{\psi}$ with
\begin{eqnarray}
\hat{\mathcal{H}}(\emph{k})=\left(
 \begin{array}{cccc}
0& \hat{h}_{1} & & \hat{h}_{_{T}}e^{-ik} \\
\hat{h}_{1}& \ddots &\!\! \ddots & \\
 &\ddots &\!\!\! \ddots &\hat{h}_{_{T-1}} \\
\hat{h}_{_{T}}e^{ik} & & \hat{h}_{_{T-1}}&\!\!\!0
 \end{array}
    \right)_{\!\!\!\!T\times T}\!\!\!\!\!\!\!\!,
    \label{K-Hamiltonian}
\end{eqnarray}
where $\hat{h}_{\alpha} = t_{\alpha}I +\lambda_{\alpha}\tau_x$ with $I$ and $\tau_{x}$ being identity matrix and the x-component of Pauli matrix acting on spin subspace, respectively, and $\alpha=1,...,T$.
When $\lambda_\alpha=\lambda_{T-\alpha}$ and $t_\alpha=t_{T-\alpha}$ , our model preserves unitary reflection symmetry about a 1D mirror point (located at the middle of unit cell) as $\mathcal{R}\hat{\mathcal{H}}(\emph{k})\mathcal{R}^{-1}= \hat{\mathcal{H}}(-\emph{k})$  with $\mathcal{R}= \delta_{i,T+1-j}\otimes\tau_x$ where $\delta_{i,j}$ is Kronecker delta. Also, because the x-component of spin is a good quantum number, the lattice has a U(1) spin rotational symmetry. So $\hat{\mathcal{H}}(\emph{k})$ is invariant under spin rotation operator $\mathcal{U} = I_T\otimes\tau_x$ around the x-axis where $I_T$ is an identity matrix of size $T$. Although the usual time-reversal symmetry is broken due to the presence of SO coupling, however, an effective time-reversal symmetry can be determined as $\mathcal{T}\hat{\mathcal{H}}(\emph{k})\mathcal{T}^{-1} = \hat{\mathcal{H}}(-\emph{k})$ with $\mathcal{T}=I_T\otimes\tau_x\mathcal{K}$ where $\mathcal{K}$ is the complex conjugate. Note the reflection, spin rotation, and time-reversal operators show the properties $\mathcal{R}^2=\mathcal{U}^2=\mathcal{T}^2=1$. Since the symmetries of Hamiltonian are based on the conventional symmetries, thus the topology of ESs can be classified following the general classification \cite{TopoClassifi,TopoClassifi1} of topological systems. Due to  $[\mathcal{R},\mathcal{T}]=0$, the topological classification of system belongs to AI class with topological index $\mathbb{Z}$.

Since the spin rotation operator commutes with $\hat{\mathcal{H}}({\emph{k}})$, the Hamiltonian can be block-diagonalized into two $T\times T$ Hamiltonians as $\hat{\mathscr{H}}({\emph{k}})=\hat{\emph{h}}_{\mathcal{U}=-}(\emph{k})\oplus\hat{\emph{h}}_{\mathcal{U}=+}(\emph{k})$ whose decoupled subspaces are spanned by eigenstates of $\mathcal{U}$ with eigenvalues $\pm1$. This can be done through a unitary transformation $\hat{\mathscr{H}}({\emph{k}})=U\hat{\mathcal{H}}(\emph{k})U^\dagger$ where $U$ is constructed from the eigenspace of $\mathcal{U}$ and will be characterized below. Each block of $\hat{\mathscr{H}}({\emph{k}})$ takes the form
\begin{eqnarray}
\hat{\emph{h}}_{\mathcal{U}=\pm}(\emph{k})=\left(
 \begin{array}{cccc}
0& \Gamma^{\pm}_{1} & & \Gamma^{\pm}_{_{T}}e^{ik} \\
\Gamma^{\pm}_{1}& \ddots &\!\!\! \ddots & \\
 &\!\!\!\ddots &\!\!\!\! \ddots & \Gamma^{\pm}_{_{T-1}}  \\
\Gamma^{\pm}_{_{T}}e^{-ik} &  & \Gamma^{\pm}_{_{T-1}}&\!\!\!0
 \end{array}
    \right)_{\!\!\!\!T\times T}\!\!\!\!\!\!\!\!,
\end{eqnarray}
where $\Gamma^{\pm}_{\alpha}=t_{\alpha}\pm\lambda_{\alpha}$. Note that $\hat{\emph{h}}_{\mathcal{U}=\pm}$ has both reflection and time-reversal symmetries because of $[\mathcal{U}, \mathcal{R}]=0$ and $[\mathcal{U}, \mathcal{T}]=0$. Now, we define the topological invariant $\mathbb{Z}$ as follows \cite{ZRefTopoInv}. Each of $\hat{\emph{h}}_{\mathcal{U}=\pm}$ commutes with reflection operator at reflection symmetric momenta $\emph{k}_{\text{ref}}=(0,\pi)$, thus eigenstates of $\hat{\emph{h}}_{\mathcal{U}=\pm}$ have a well-defined parity $\mathcal{\zeta}_{\mathcal{U}=-(+)}(\emph{k}_{\text{ref}})=\pm1$ at those points. This, subsequently, allows for specifying an integer invariant $\mathcal{N}_{i,\mathcal{U}=\pm}=|\emph{n}_{1,i,\mathcal{U}=\pm}-\emph{n}_{2,i,\mathcal{U}=\pm}|$ to classify $\hat{\emph{h}}_{\mathcal{U}=\pm}$. Here, we have defined $\emph{n}_{1,i,\mathcal{U}=\pm}$ and $\emph{n}_{2,i,\mathcal{U}=\pm}$ as the number of negative parities related to the energy bands of $\hat{\emph{h}}_{\mathcal{U}=\pm}(\emph{k})$ in the $\textit{i}$th bandgap at $\emph{k}_{\text{ref}}=0$ and $\emph{k}_{\text{ref}}=\pi$, respectively. So, topological number $\mathbb{Z}$ for multi-subspace and multi-band structure of the system can be defined as
\begin{eqnarray}
\mathbb{Z}:=\sum\limits_{\emph{j}=\pm}Z_{\mathcal{U}=j}=\sum\limits_{\emph{j}=\pm}\sum\limits_{\emph{i}=1}^{T-1}\mathcal{N}_{i,\mathcal{U}=j},
\label{Z index}
\end{eqnarray}
giving the number of localized ESs under open boundary conditions. Here, $Z_{\mathcal{U}=\pm}$ denotes the topological invariant of subspaces.

Interestingly, each subsystem, described by $\hat{\emph{h}}_{\mathcal{U}=\pm}$, is similar to a 1D spinless system consisting of $T$ "super-sublattices" per unit cell. Each super-sublattice is comprised of a sublattice with opposite spin species so that the new hopping amplitude between two adjacent super-sublattices in the subsystem labelled by $\mathcal{U}=\pm$ is $\Gamma^{\pm}_{\alpha}$. This can be illuminated by a transformation from the old basis to the new one through the unitary matrix $U_{2T\times2T}$ as $\hat{\Psi}_i=\sum_{j=1}^{2T}U_{i,j}\hat{\psi}_{j}$ where $i=1,...,2T$. The non-zero matrix elements of $U$ are
\begin{align}
U_{\alpha, 2(T-\alpha+1)}&=\frac{1}{\sqrt{2}},&\quad U_{2T-\alpha+1, 2\alpha-1}&=\frac{1}{\sqrt{2}},&\nonumber\\
U_{\alpha, 2(T-\alpha)+1}&=\frac{-1}{\sqrt{2}},&\quad U_{2T-\alpha+1, 2\alpha}&=\frac{1}{\sqrt{2}}.\nonumber&
 \end{align}
Therefore, the new basis is $\hat{\Psi}=(\hat{\Psi}^{-},\hat{\Psi}^{+})^\textit{\textbf{\!T}}$ where $\hat{\Psi}^{-(+)}$ corresponds to the basis of eigenspace of $\mathcal{U}$ with eigenvalues $-1(+1)$ whose entries are super-sublattices given by $\hat{\Psi}_\alpha^{\pm}=1/\sqrt{2}(\hat{c}_{T-\alpha+1,k,\downarrow}\pm\hat{c}_{T-\alpha+1,k,\uparrow})$.

On the other hand, it is also easy to obtain real-space operator for the spin rotational symmetry as $\mathscr{U}=I_{T N}\otimes \tau_x$ where $I_{T N}$ is an identity matrix of size $TN$ with $N$ being the number of unit cells. Therefore, the real-space Hamiltonian (\ref{hamiltonian-real space-total}) can be brought into two block-diagonal matrices in the eigenspace of $\mathscr{U}=\pm$ as $\hat{\mathfrak{H}}=\hat{\mathfrak{h}}_{\mathscr{U}=-}\oplus\hat{\mathfrak{h}}_{\mathscr{U}=+}$. In order to study the localization of states, we calculate normalized logarithm of the inverse participation ratio (IPR) of an eigenvector $|\psi_E\rangle$ associated with eigenenergy $E$ as defined by $I_E=\text{Ln}(\sum_{i=1}^{2TN}|\langle i|\psi_E\rangle|^4)/\text{Ln}(2TN)$ where $|i\rangle$ is basis elements \cite{IPR}. Here, $I_E=-1$  denotes delocalized states, whereas for much more localized ones $I_E=0$.\\

{\it Results and discussion.}---Without loss of generality, we focus on the case of three sublattices per unit cell, $T=3$, as the model illustrated in Fig. \ref{Fig1}(a). Reflection symmetry requires $\lambda_1=\lambda_2\equiv\lambda^\prime$ and $t_1=t_2\equiv t^\prime$. Here, the intra and inter unit cell hoppings, respectively, are $t^\prime=t-\delta t$ and $t_3=t+\delta t$ where $t$ ($\delta t$) stands for hopping energy (hopping modulation strength). Energy spectrum of $\hat{\mathfrak{H}}$ and its IPR as a function of intra unit cell SO coupling strength $\lambda^\prime$ under open boundary conditions is shown in Fig. \ref{Fig1}(b) with inter unit cell SO coupling strength $\lambda_3=-\lambda^\prime$ and $\delta t<0$. The solid lines and hexagrams represent the eigenvalues of $\hat{\mathfrak{h}}_{\mathscr{U}=-}$ and $\hat{\mathfrak{h}}_{\mathscr{U}=+}$, respectively. As $\lambda^\prime$ increasing, two gap closings occur simultaneously in the subbands of $\hat{\mathfrak{h}}_{\mathscr{U}=-}$ away from Fermi energy at $\lambda^\prime=0.5t$ and then two degenerate localized ESs emerge in the bandgaps. Interestingly, with further increase in $\lambda^\prime$, the pairs of ESs enter to the GBSs of $\hat{\mathfrak{h}}_{\mathscr{U}=+}$ at $\lambda^\prime=t$. These topological ESs meet each other at Fermi energy with $\lambda^\prime=t^\prime=1.5t$ leading to the appearance of strongly localized fourfold degenerate states. Finally, they stay in the GBSs while preserving their localization with an extra increase of $\lambda^\prime$. The corresponding topological $\mathbb{Z}$ integer is plotted in Fig. \ref{Fig1}(d). In the parameter region where the ESs appear, $\mathbb{Z}$ integer takes value 2 demonstrating the existence of two pairs of localized ESs steaming from $Z_{\mathcal{U}=-}=2$ (shown in Fig. \ref{Fig1}(c)) whereas $Z_{\mathcal{U}=+}=0$.
As a result, topologically protected ESs of an eigenspace could penetrate into trivial GBSs of the other one.

In particular, both of the eigenspaces may host nontrivial topological phases by choosing appropriate SO coupling values in a way that topological phase transitions occur in both subspaces. This will result in appearance of four pairs of ESs. In Figs. \ref{Fig1}(e) and \ref{Fig1}(g), respectively, the dependence of energy spectrum $\hat{\mathfrak{H}}$ and topological invariant $\mathbb{Z}$ on $\lambda^\prime$ is presented for $\delta t>0$ and $\lambda_3=\lambda^\prime+2.5t$. As shown in Figs. \ref{Fig1}(e) and \ref{Fig1}(f), the first topological phase transition happens in the $\hat{\mathfrak{h}}_{\mathscr{U}=+}$ spectrum leading to appearance of highly localized ESs in the bandgaps (GBSs) for the parameter space $\lambda^\prime\in (-2.25t,-0.75t)[(-0.75t,-0.05t)]$. After taking place of the second topological phase transition in the $\hat{\mathfrak{h}}_{\mathscr{U}=-}$ spectrum at $\lambda^\prime=-0.25t$, two new ESs are emerged in addition to the former ones with $Z_{\mathcal{U}=-}=2$. Therefore, the system hosts four topological ESs $\mathbb{Z}=4$, as shown in Fig. \ref{Fig1}(g). When $\lambda^\prime$ further increases, surprisingly, topological ESs of the $\hat{\mathfrak{h}}_{\mathscr{U}=-}$ spectrum reside inside the nontrivial GBSs of $\hat{\mathfrak{h}}_{\mathscr{U}=+}$ and system re-enters to the TM phase. This is in contrast to the case of topological bound states embedded in non-topological continuous spectrum \cite{BIC}. Moreover, the electron-like and hole-like ESs intersect each other at $\lambda^\prime=\pm t^\prime=\pm0.5t$ [Fig. \ref{Fig1}(e)]. Remarkably, from both Figs. \ref{Fig1}(b) and \ref{Fig1}(e), one can see that ESs of an eigenspace at Fermi energy are quasi-degenerate not only with their own highly degenerate GBSs but also with GBSs of the other eigenspace establishing TM phase. Note also that the $\mathbb{Z}$ values change at which the subband gap closing/reopening occurs [Figs. \ref{Fig1}(d) and \ref{Fig1}(g)].

In order to shed light on the mechanisms underlying the above-mentioned behaviors, we focus on understanding the effect of spin and sublattice degrees of freedom on band structure. In fact, the odd number of sublattices provides bulk metallic ground states resulting in breaking particle-hole and chiral symmetries. Therefore, possible bandgaps can only occur away from Fermi surface. Now, exploiting SO coupling breaks spin degeneracy and subsequently each band splits into two subbands corresponding to two different helical components. The resulting spin helical subbands retain the nontrivial ESs and at the same time push them into the metallic bulk states as a consequence of U(1) symmetry. As such, if the system is in a topologically nontrivial phase then the Fermi level crosses at some of the topological ESs embedded in GBSs.

The topological phase diagram of system in the plane ($\lambda^\prime,\lambda_3$) is depicted in Figs \ref{Fig2}(a) and \ref{Fig2}(b) for $\delta t = -t/2$ and $\delta t = +t/2$, respectively. The black solid lines denote the border between topologically different phases. Also, the dashed lines correspond to TM states at Fermi level. The topological phase diagrams are categorized into three distinct phases according to $\mathbb{Z}=0, 2, \text{and}\ 4$ implying none, two, and four pairs of ESs, respectively. The regions $\mathbb{Z}= 2\ \text{and}\ 4$ are divided into subcategories depending on the appearance of ESs either within GBSs or in the bandgaps. In both diagrams, there are three possibilities for the region $\mathbb{Z}=4$: i) four pairs of ESs within GBSs, ii) four pairs of ESs in the bandgaps, and iii) two pairs of ESs in the bandgaps and the other two pairs of ESs within GBSs. In addition, the regions $\mathbb{Z}=2$ have two possibilities that are two pairs of ESs within GBSs and within bandgaps. Also, the regions of $\mathbb{Z}=2$ and $\mathbb{Z}=4$ with ESs in the bandgaps for $\delta t = +t/2$ are more dominant than those of $\delta t = -t/2$.

{\it Stability of ESs.}---The existence of topological ESs that are quasi-degenerate with GBSs is ensured by the presence of spin rotational symmetry. To illustrate this feature, let us investigate ESs stability against perturbations originated from on-site potential and Zeeman magnetic field. We add the term $\hat{H}^\prime=\hat{H}_{V}+\hat{H}_{B}$ to Hamiltonian (\ref{hamiltonian-real space-total}) including on-site Hamiltonian $\hat{H}_{V}=\sum_{n,\sigma}\sum_{\alpha}^{T}V_{\alpha,n}\hat{c}^\dagger_{\alpha,n,\sigma}\hat{c}_{\alpha,n,\sigma}$ and Zeeman Hamiltonian $\hat{H}_{B}=\sum_{n,\sigma,\sigma^{\prime}}\sum_{\alpha}^{T}\hat{c}^\dagger_{\alpha,n,\sigma}(\mathbf{M}_{n,\alpha}\cdot\boldsymbol{\tau})\hat{c}_{\alpha,n,\sigma^{\prime}}$. Here, $V_{\alpha,n}$ defines the amplitude of on-site potential, $\mathbf{\tau}$ is the Pauli spin vector and the Zeeman field vector is $\mathbf{M}_{n,\alpha}=(M_{n,x,\alpha},M_{n,y,\alpha},M_{n,z,\alpha})$. For concreteness, we will inspect the effects of these perturbations on the topological properties separately.

\begin{figure}[t]
  \centering
  \includegraphics[width=8.5cm]{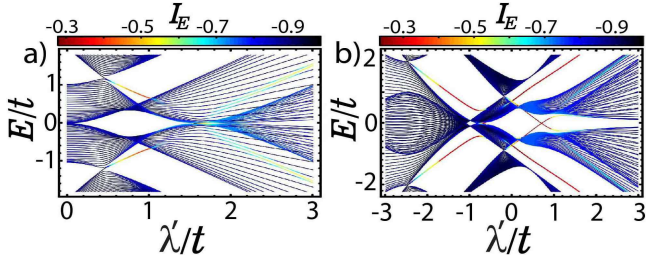}
    \caption{(Color online) Band structure with its IPR as a function of $\lambda^\prime$ for (a) $\delta t = -t/2$, (b) $\delta t = +t/2$ in the presence of y-component of staggered Zeeman field chosen as $M_{y,1}=M_{y,3}$ and $M_{y,1}=-M_{y,2}=t/2$. Other parameters of panels (a) and (b) are the same as Figs. \ref{Fig1}(b) and \ref{Fig1}(e), respectively.}
   \label{Fig3}
\end{figure}

We first investigate only the effect of Zeeman field. Since the lattice is invariant under rotations about the x-axis, we apply Zeeman field along the y-axis. This Zeeman field violates lattice U(1) symmetry and, in consequence, $\hat{\mathcal{H}}({\emph{k}})$ can not be block diagonalized. Accordingly, the topological invariant (\ref{Z index}) is no longer valid. To discriminate the role of U(1) symmetry from that of reflection symmetry, we break U(1) symmetry such that the reflection symmetry remains untouched. To do so, we need to set $M_{y,1}=M_{y,3}$. Under such situation, the energy spectrum of Figs. \ref{Fig1}(b) and \ref{Fig1}(e) is re-calculated and plotted in Figs. \ref{Fig3}(a) and \ref{Fig3}(b), respectively, in the presence of y-component of staggered Zeeman field. Interestingly, one can observe that ESs within GBS can not survive resulting in termination of ESs from at least one end with no gap closure. While in-gap states preserve their degeneracies owing to the reflection symmetry. Although this feature is obtained for staggered Zeeman field but similar results hold for uniform Zeeman field. Otherwise, preserving the spin rotational symmetry and violating reflection one lead to destroying ESs and making the system topologically trivial (not shown). As a result, the TM phase manifests itself whenever both reflection and U(1) symmetries are present simultaneously in the system.

In addition, the x-component of Zeeman field or on-site potential conserve the U(1) symmetry. One readily finds that these terms have diagonal entries in each block of transformed Hamiltonian $\hat{\mathscr{H}}({\emph{k}})=(\hat{\emph{h}}_{\mathcal{U}=-}(\emph{k})+\hat{\mathcal{V}}_{\mathcal{U}=-})\oplus(\hat{\emph{h}}_{\mathcal{U}=+}(\emph{k})+\hat{\mathcal{V}}_{\mathcal{U}=+})$ with $\hat{\mathcal{V}}_{\mathcal{U}=\pm}=\mu^{\pm}_{i}\delta_{i,j}$ where $\mu^{\pm}_{i}=V_{T-i+1}\pm M_{x,T-i+1}$. Here, reflection symmetry requires $V_i=V_{T-i+1}$ and $M_{x,i}=M_{x,T-i+1}$, for which we set $V_1=V_3$ and $M_{x,1}=M_{x,3}$ in the case of $T=3$. Obviously, the uniform on-site potential (x-component of Zeeman field) shifts the energy levels of each block of $\hat{\mathscr{H}}({\emph{k}})$ to the same (opposite) direction. These enable us to shift the energies of ESs appearing away from Fermi level toward $E=0$ while their quasi-degeneracy with GBSs remain intact. Therefore, the TM phase will be accessible for much larger range of SO couplings. Moreover, interestingly, either alternating on-site potential or x-component of Zeeman splitting can impose gap closing in blocks of $\hat{\mathscr{H}}({\emph{k}})$ independently. So, the number of ESs would be asymmetric about Fermi level resulting in inducing odd number of ESs.
\begin{figure}[t]
  \centering
  \includegraphics[width=8.5cm]{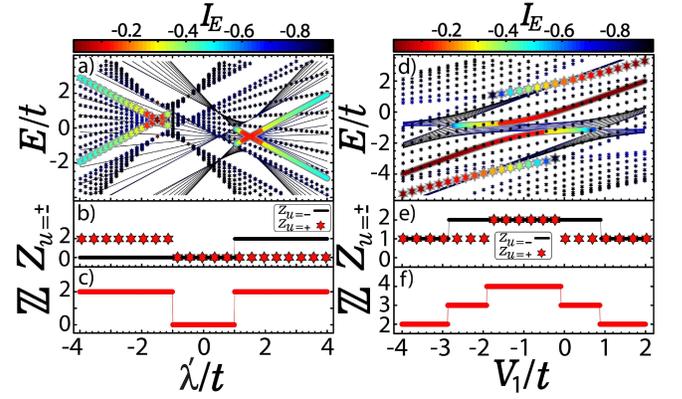}
    \caption{(Color online) Energy spectrum of $\hat{\mathfrak{H}}$ with its IPR versus (a)[(d)] $\lambda^\prime$ ($V_1$) with $\delta t=-t/2$: $\lambda_3=\lambda^\prime$ and $M_{x,1}=M_{x,2}=M_{x,3}=t/2$  ($\lambda^\prime=t,\ \lambda_3=2.5t,\ V_3=V_1$ and $V_2=-t$) for 60 unit cells. Panels (c) and (f) [(b) and (e)] are the corresponding invariant $ \mathbb{Z}\ (Z_{\mathcal{U}=\pm})$.}
   \label{Fig4}
\end{figure}
The energy spectra of a finite chain as functions of $\lambda^\prime$ in the presence of a uniform Zeeman field along x-axis and of $V_1$ are depicted in Figs. \ref{Fig4}(a) and \ref{Fig4}(d), respectively. One can see from Fig. \ref{Fig4}(a) that the x-component of Zeeman field splits the energy spectra of $\hat{\mathfrak{h}}_{\mathscr{U}=-}$ and $\hat{\mathfrak{h}}_{\mathscr{U}=+}$ relative to each other, as already mentioned. Also, three pairs of ESs can be seen in the energy spectrum of Fig. \ref{Fig4}(d). The corresponding invariants $\mathbb{Z}$ ($Z_{\mathcal{U}=\pm}$) are shown in panels (c) and (f) [(b) and (e)] of Fig. \ref{Fig4}.

{\it Experimental proposal.}---Recent experimental achievements make it possible to realize 1D dimerized lattice model, known as Su-Schrieffer-Heeger (SSH) model \cite{1DTI}, by fabricating heterostructures of alternating thin films of band insulators and TIs \cite{SSHCondMatt} relying on condensed matter physics. Also, modulated SO coupling can be implemented either by applying local electric field \cite{Rashba} or by using cluster of heavy atoms \cite{SOProxy}. In the latter case, due to proximity effect, SO coupling can be transferred from the bands of heavy atoms to the bands of system such that other properties of the structure itself remain unaffected. On the other hand, using cold atoms in optical lattices provide an excellent playground with the easy tunability of control parameters to simulate topological bands \cite{TopoBandColdAtom} in artificial quantum systems like SSH chain \cite{SSHExperCold}. For cold-atom experiments, we suggest to employ superposition of retroreflected laser beams or to imprint superlattice with a spatial light modulator producing extended SSH model with three number of sublattices \cite{Tripartit}. Furthermore, it is possible to engineer SO interaction in tripartite lattices \cite{SOTripartit} even for neutral cold-atomic gases \cite{SOcoldAtoms}. Using spatially resolved radio-frequency spectroscopy \cite{RadioSpectr}, the ESs within GBSs can be recognized from the local density of states.

{\it Conclusions.}---We revealed a new kind of exotic topological states characterized by the coexistence of topologically nontrivial ESs and nontrivial GBSs either at or far from the Fermi level. The main ingredient of these exotic metallic states arises from the coupling of odd number of sublattices to spin degree of freedom in a 1D periodic arrays. We found that such systems undergo a topological phase transition under subband gap closure conditions. The effects of Zeeman fields and on-site potentials on the topological phases indicate that TM states are protected by both reflection and spin rotational symmetries. The concept of nontrivial TM phase may be generalized to higher invariants and non-hermitian case, as well as including interaction effects.

\section*{Acknowledgment}
This work is partially supported by Iran Science Elites Federation under Grant No. 11/66332.

\end{document}